\documentclass[aps,pra,twocolumn,groupedaddress]{revtex4-1}

\usepackage{amsmath}
\usepackage{amsthm}
\usepackage{amssymb}
\usepackage{graphicx,hyperref,bbm,enumitem,bm}  
\usepackage[pdftex]{color}

\usepackage[english]{babel}

\usepackage{xcolor}

\usepackage{float}

\usepackage[toc,page]{appendix}

\def\1{\mathbbm{1}}
\def\ket#1{{| #1 \rangle}}
\def\bra#1{{\langle #1 |}}
\def\rA{\rho_{\rm A}}

\def\leff{\lambda_{\mathrm{eff}}}
\def\ceff{c_{\mathrm{eff}}}
\def\cbw{c_{\mathrm{BW}}}
\def\cs{c_{\mathrm{S}}}

\def\tit#1{{\em #1},}

\newcommand{\new}[1]{{\color{black} #1}}

\begin{document}
	
	\title{Purity decay rate in random circuits with different configurations of gates}
	
	\author{Ja\v s Bensa and Marko \v Znidari\v c}
	\affiliation{Department of Physics, Faculty of Mathematics and Physics, University of Ljubljana, 1000 Ljubljana, Slovenia}
	
	\date{\today}
	
	\begin{abstract}
          We study purity decay -- a measure of bipartite entanglement -- in a chain of $n$ qubits under the action of various geometries of nearest-neighbor random two-site unitary gates. We use a Markov chain description of average purity evolution, using further reduction to obtain a transfer matrix of only polynomial dimension in $n$. In most circuits, an exception being the brick-wall configuration, purity decays to its asymptotic value in two stages: the initial thermodynamically relevant decay persisting up to extensive times is $\sim \leff^t$, with $\leff$ not necessarily being in the spectrum of the transfer matrix, while the ultimate asymptotic decay is given by the second largest eigenvalue $\lambda_2$ of the transfer matrix. The effective rate $\leff$ depends on the location of bipartition boundaries as well as on the geometry of applied gates.
	\end{abstract}
	
	\maketitle
	
\section{Introduction}

Entanglement \cite{entanglement} is considered one of the characteristic traits of quantum mechanics \cite{schrodinger}, and is responsible for some seemingly counterintuitive predictions \cite{EPR}. As years passed, entanglement went from being a subject of philosophical debates to a quantum resource \cite{QIT,cryptography,feynman}. Studies of entangled systems became a central aspect of research in quantum information. Understanding entanglement growth in chaotic many-body systems is important because complex systems may manifest new phenomena not present in simpler systems \cite{anderson}. Moreover, chaotic systems may be used to generate entanglement quickly and robustly \cite{chaos_book,page,page_proof}. Unfortunately, analytical solutions of complex systems are usually impossible to obtain, so solvable toy-models are often utilized to reproduce certain properties of more complex many-body systems.

Random quantum circuits \cite{emerson03,RQC_review} are a widely used toy-model used to describe certain properties of chaotic systems\cite{Chalker18,AdamPRB19}, or, ultimately, that of completely random unitaries as in $t$-designs \cite{gross07,Harrow09,brandao16,brandao16b,Hunter20}. Regarding entanglement, random quantum circuits allowed for many analytical results or simplifications \cite{Zanardi_12,HJ_PRA22,e_membrane}; some examples are Markov chain reductions \cite{oliveira,markov,PRA08,PRX,marko_22,you21}, statistical mechanical descriptions \cite{adam18,AdamPRB19,vasseur21,altman20} and solvable results in dual-unitary circuits \cite{sarang19,du19,Bruno20}. In this article, we explore how average bipartite entanglement of a qubit chain evolves in different random quantum circuits composed of random two-site gates. Previous research mostly focused on the so called brickwall (BW) \cite{Frank18,adam18,AdamPRB19,PRX17,markov,PRX}, and more recently also staircase (S) \cite{marko_22,adam18,PRX} circuits (see Fig.~\ref{fig:BW_and_S} for illustration). For a single-cut bipartition and open boundary conditions it is known that in a brick-wall circuit composed of Haar random gates purity decays towards its asymptotic value as $\sim(4/5)^{2t}$\cite{markov,Frank18,adam18,AdamPRB19,u4_conjecture}, whereas it goes as $(2/3)^t$ for a staircase configuration of gates~\cite{PRX,marko_22}.  We shall extend those results to more general circuit geometries and bipartitions, finding that the entanglement growth, i.e., purity decay, can be much richer and is essentially given by a product of the above two factors.

We work with the simplest entanglement quantifier, purity $I_{A}(t) = {\rm tr}_{\rm A}\rA^2(t)$. As the entanglement in the system grows in time, purity decays to its asymptotic value $I(\infty)$. Using purity instead of some other quantity simplifies analytical and numerical calculations. Namely, purity, averaged over all possible choices of the random gates in the quantum circuit, can be evolved using a Markov chain \cite{markov,oliveira,PRX17}. With the help of this Markov chain purity calculation reduces to an iteration of a transfer matrix on a vector. If we calculate the transfer matrix following the procedure from \cite{PRA08,markov}, the dimension of the transfer matrix is $2^n \times 2^n$, where $n$ is the number of qubits, however, it was recently shown \cite{marko_22} that, for S circuits, the dimension can be reduced to $n \times n$, greatly simplifying analysis. We extend this reduction method to other circuit geometries.

\begin{figure}[h]
    \begin{center}
        \includegraphics[width=80mm]{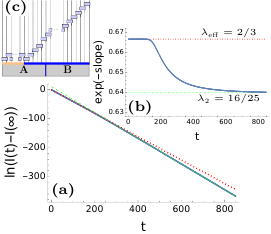}
        \caption{Decay of purity for a bipartition separating the first half of qubits from the rest in a random quantum circuit composed of a brick-wall (BW) like section acting on the first $50$ qubits and a staircase (S) section acting on the remaining $150$, $n=200$ and open boundary conditions (OBC). The circuit protocol used is schematically depicted in the panel (c), where the orange \new{(light grey)} line denotes the extend of the BW part of the circuit. Meanwhile the blue \new{(dark grey)} line shows the region under the S part. The picture (b) shows the instantaneous decay rates of purity, the exponent of the slopes of lines in panel (a), compared with $\leff$ (red dotted line) and $\lambda_2$ (green dashed line). The decay to $I(\infty)$ is $(2/3)^t$ (red dotted line) until the time $t\approx n$; afterwards, it is given by $|\lambda_2| = 16/25$ (green dashed line).}
        \label{fig:phantom_example}
    \end{center}
\end{figure}

Recently it was noted that purity dynamics under the action of random quantum circuits can exhibit counterintuitive features \cite{PRX,marko_22}. Namely, up to times extensive in the number of qubits, $t \sim n$, the exponential purity decay to $I(\infty)$ is not always determined by the second largest eigenvalue $\lambda_2$ of the Markov chain transfer matrix (note that the largest eigenvalue is equal to $1$) as one could expect. Purity instead can decay to $I(\infty)$ as

\begin{equation}
    |I_A(t)-I(\infty)| \sim \leff^t,
\label{eq:I_decay}
\end{equation}
where the effective decay $\leff$ is not equal to $\lambda_2$. In fact, it can be either smaller or, surprisingly, greater than $\lambda_2$ -- in such a case $\leff$ has been dubbed~\cite{PRX} a phantom ``eigenvalue'' because it is not in the spectrum of any finite transfer matrix (see Fig.~\ref{fig:phantom_example} for an example). The peculiarity of this phenomenon lies in the fact that the behaviour determined by $\leff$ persists up to times $\sim n$, which makes $\leff$ the correct and relevant decay in the thermodynamic limit (TDL) $n \rightarrow \infty$. For S circuits with Haar random gates such phantom decay has been numerically observed in Ref.~\cite{PRX} and analytically explained in Ref.~\cite{marko_22}. The discrepancy between $\leff$ and $\lambda_2$ was also observed in out-of-time-order correlators (OTOC) \cite{PRR}, which are more easily measured in experiments, see Ref.~\cite{google_otoc} for a measurement in a random circuit. The reason for the discrepancy between $\lambda_2$ and the true decay $\leff$ still needs to be fully understood. It is worth mentioning, though, that recently a number of phenomena has been found whose origin can similarly be traced back to odd effects that non-Hermiticity can have, and that perhaps have a related origin. Examples are relaxation rates under Lindblad evolution~\cite{Mori20,ueda21,wang19,clerk22,yang22}, nondiagonalizability and Jordan blocks~\cite{foot0} in, e.g., integrable circuits \cite{lamacraft21}.

To understand how is it possible that the asymptotic decay is not given by $\lambda_2$, one must look at the spectral decomposition of the transfer matrix. In our case, the transfer matrix is non-Hermitian, so its right (and left) eigenvectors are not orthogonal between each other; the only condition one has is biorthogonality $ \langle l_k|r_j\rangle = \delta_{j,k}$. If we choose to normalize all right eigenvectors $ |r_k\rangle$ to $1$, then the norm of left eigenvectors $ \langle l_k|$ can grow exponentially large with $n$ (this can be seen explicitly for the exact solution in Ref.~\cite{marko_22}), delaying the appearance of $\lambda_2^t$ to extensive times. See also \cite{mori21,sarang21,viola21} for related phenomena caused by an explosion of coefficients. If the transfer matrix would be Hermitian, the triangle inequality would bound the expansion coefficients and the asymptotic decay would be given by $\lambda_2$, therefore the effect could not happen.  

Studying different circuits and different bipartitions we find that one has $\leff \neq \lambda_2$ rather generically, and not just for the S configuration. The BW circuit where $\leff=\lambda_2$ is therefore special. Based on our results we make a conjecture that $\leff$ depends on the number of bipartition boundaries that fall in the S and in the BW section of a circuit.

Section~\ref{sec:II} of the paper is dedicated to the formalism that we use to describe average purity in random quantum circuits. In section~\ref{sec:IIA} we give a description of the circuits that we use and talk about the Markov chain description of purity evolution. In the section~\ref{sec:IIB} we describe the method from \cite{marko_22} used to reduce the dimensionality of the problem. In Sec.~\ref{sec:III} we shall analyze purity for bipartitions that cut the first $k$ qubits from the others, i.e., single-cut bipartitions. We derive the reduced transfer matrix for different random quantum circuits and calculate the effective decay $\leff$. Lastly, in sec.~\ref{sec:IV} we generalize the results obtained in sec.~\ref{sec:III} to arbitrary bipartitions.

%\section{Formalism and previous results}
\section{Reduced transfer matrix description of average purity}
\label{sec:II}

\subsection{Purity in random quantum circuits}
\label{sec:IIA}

\begin{figure}[h]
    \begin{center}
        \includegraphics[width=80mm]{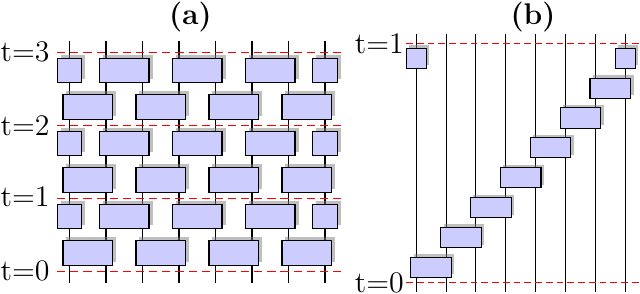}
        \caption{(a) The brick-wall (BW) protocol and (b) the staircase (S) protocol on a qubit chain with $n=8$ qubits and periodic boundary conditions (PBC). Blue boxes represent independent random two-site gates $U_{i,i+1}$, and red dotted lines mark integer times.}%
        \label{fig:BW_and_S}
    \end{center}
\end{figure}

The random quantum circuits that we focus on are defined as a product of independent identically distributed Haar nearest neighbour two-site gates $U_{i,i+1}$ (in this notation $U_{i,i+1}$ couples the qubits $i$ and $i+1$). By Haar gates we mean unitary operators distributed according to the unique unitary invariant Haar measure on U(4) \cite{Haar}. When dealing with qubits on a chain with open boundary conditions (OBC), the unitary propagator $U$ for a single time step will consist of $n-1$ different gates (acting on different nearest neighbour qubit pairs). Similarly, in the case of periodic boundary conditions (PBC) the total number of gates in a single time step is $n$. In both cases one can vary the order of the two qubit gates in $U$, thereby obtaining a factorial number of different protocols (i.e., configurations or geometries). The most studied configuration of two-site gates is the brick-wall (BW) configuration (Fig.~\ref{fig:BW_and_S}(a)), while the staircase (S) configuration (Fig.~\ref{fig:BW_and_S}(b)) also plays a prominent role. In this work we also study other configurations, showing that changing the configuration and the bipartition can greatly influence the decay of purity.

Entanglement will be quantified using purity. Let $A$ and $B$ be two complementary subsets of the set of our $n$ qubits, namely $A \cup B = \{ 1, \dots, n \}$. Purity for the bipartition $A/B$ is defined as
\begin{equation}
     I_{A}(t) = {\rm tr}_{\rm A}\rA^2(t),
\label{eq:purity_def}
\end{equation}
where $\rA(t)={\rm tr}_{\rm B} \ket{\psi(t)}\bra{\psi(t)}$ and ${\rm tr}_{\rm A}$ denotes the partial trace over the subset $A$. The evolved pure state $\ket{\psi(t)}$ at integer $t$ is obtained simply as $\ket{\psi(t)} = U^t \ket{\psi(t=0)}$. Each application of $U$ evolves the state by one unit of time, and even though we used a short notation $U^t$ this does not mean that we repeat the same gates $U_{i,i+1}$ in every period -- all two-qubit gates $U_{i,i+1}$ at each time step are independent Haar random.

We are interested in how purity of an initially separable state, which has $I_A(t=0)=1$, evolves towards its long-time asymptotic value $I(\infty)$. Under the action of random quantum circuits one has $I(\infty) = (2^{|A|}+2^{|B|})/(1+2^n)$ ($|C|$ denotes the number of elements of the set $C$) \cite{Lubkin}. Note that a decrease in purity means an increase in bipartite entanglement. Averaging over random gates $U_{i,i+1}$ composing our random quantum circuit, it is possible to map the evolution of average purity to a Markov chain \cite{oliveira,markov}. Following the procedure explained in \cite{markov} we end up with a transfer matrix $M$ propagating a vector $\vec{I}(t)$, whose coefficients are purities of a state $ |\psi(t) \rangle$  for all $2^n$ possible bipartitions of $n$ qubits. Writing $\vec{I}=([\vec{I}]_0,[\vec{I}]_1,\dots,[\vec{I}]_{2^n-1})$, the ``bipartition'' index $\alpha$, i.e., $[\vec{I}]_\alpha$, encodes the bipartition by prescription $\alpha = \sum_{j \in A} 2^{j-1}$. In other words, if $\alpha$ is written in a binary representation, its bits that are set to $1$ mark qubits that are in the subsystem $A$. For instance, for $n=8$ and the bipartition $A=\{ 1,2,5,7\}$ the corresponding bitstring is $11001010$, i.e., decimal \new{$\alpha=83$} (bits are ordered from left to right, for qubits $1$ to $n$). To obtain average purities at time $t+1$ from purities at time $t$ we simply have to multiply $\vec{I}(t)$ by a matrix $M$,
\begin{equation}
    \vec{I}(t+1) = M \vec{I}(t).
    \label{eq:Markov_chain}
\end{equation}
In this paper we will not explain how to obtain $M$, we instead just state the result~\cite{PRA08,markov,PRX17}. Similarly as $U$, $M$ is also a product of two-site gates $M_{i,i+1}$ multiplied in the same order (protocol) as gates in $U$. For example, for the random quantum circuit in the BW configuration (Fig.~\ref{fig:BW_and_S}(a)) on $8$ qubits with PBC we have $U = U_{8,1} U_{6,7} U_{4,5} U_{2,3} U_{7,8} U_{5,6} U_{3,4} U_{1,2}$, and therefore
\begin{equation}
    M = M_{8,1} M_{6,7} M_{4,5} M_{2,3} M_{7,8} M_{5,6} M_{3,4} M_{1,2},
\label{eq:M_BW_PBC_n8}
\end{equation}
with the two-site matrix $M_{i,i+1}$ being for our Haar random two-qubit gate
\begin{equation}
    M_{i,i+1} = \begin{pmatrix}
                    1 & 0 & 0 & 0         \\
                    a & 0 & 0 & a       \\
                    a & 0 & 0 & a       \\
                    0 & 0 & 0 & 1
              \end{pmatrix},
    \label{eq:Mij}
\end{equation}
where $a = 2/5$ for qubits, however, Eq.~\ref{eq:Mij} works also for arbitrary qudits, i.e., local Hilbert space of dimension $d$ with $a = d/(d^2+1)$. All our results presented in the following sections therefore hold for any $d$, although numerical results will be presented for qubits.

For an initial state that is separable with respect to any bipartition, i.e., a product state of single-qubit states, the initial vector of purities is $\vec{I}(t=0)=(1,1,\dots,1)$. To get purities at a later time one must propagate it with the $2^n \times 2^n$ matrix $M$. Due to an exponentially large dimension, however, such iteration is not efficient. In the next section we explore a new method that in certain cases allows us to reduce the dimension of the transfer matrix $M$ from $2^n$ down to a dimension that is only linear in $n$.

\subsection{Reduction of the transfer matrix}
\label{sec:IIB}

In this section we describe the procedure found in Ref.~\cite{marko_22} to reduce the relevant dimension of the transfer matrix from $2^n$ to just $\sim n$ in the case of the S or BW configurations with OBC. It was observed that, for a single-cut bipartition, i.e., a bipartition where the first $k$ qubits are in $A$ and all others are in $B$, one must keep track of only $n$ other single-cut bipartition purities instead of all $2^n$ purities -- correspondingly, the relevant reduced transfer matrix for single-cut bipartitions and the S (or BW) configuration of gates was of dimension $n \times n$.

\begin{figure}[h]
    \begin{center}
        \includegraphics[width=80mm]{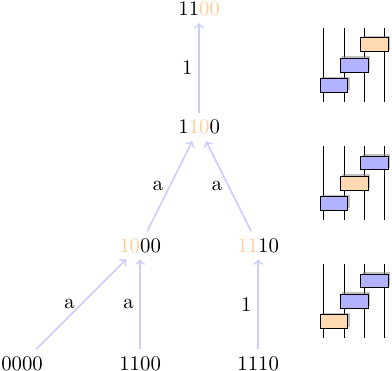}
        \caption{Illustrating the propagation of single-cut purities $I_k(t-1)$ on $n=4$ qubits and the S protocol, $M=M_{3,4} M_{2,3} M_{1,2}$. Shown are contributions to $I_2(t)$ (top bitstring). At each node of the graph we apply one two-site $M_{i,i+1}$ (\new{orange/light grey} gate in circuits on the right acting on two bits highlighted in \new{orange/light grey}), beginning with $M_{3,4}$ and sequentially moving down to $M_{1,2}$ (read from top to bottom). The arrows connect bitstrings that contribute to them after gate application, with the coefficients written at a side. For more details see the main text.}
        \label{fig:basis_example}
    \end{center}
\end{figure}

To simplify notation we denote by $I_k(t)$ purity at time $t$ for the single-cut bipartition in which the first $k$ qubits are in $A$, the other in $B$ (note that in previous section we used $[\vec{I}]_{\alpha}$ to denote a specific component -- a generic bipartition $\alpha$, now we use $I_k$ to denote a single-cut bipartition, i.e., in $I_k = [\vec{I}]_{\alpha=2^k-1}$). We want to compute $I_k(t)$, $k=1,\ldots,n-1$ and we analyze which bipartitions at the previous time $t-1$ can contribute to $I_k(t)$. It is convenient to work with the bitstring representation of bipartitions. In the bitstring notation the basis for the matrix $M_{i,i+1}$ from Eq.~\ref{eq:Mij} is $\{00,10,01,11\}$, so $M_{i,i+1}$ maps the bits at positions $i$ and $i+1$ in the bitstring as
\begin{align}
    &00 \xleftarrow{1} 00, \nonumber \\
    &01 \xleftarrow{a} 00,11, \nonumber \\
    &10 \xleftarrow{a} 00,11, \nonumber \\
    &11 \xleftarrow{1} 11, 
\label{eq:maps}
\end{align}
meaning that we can get $00$ only from $00$, $11$ from $11$, and $01$ or $10$ from either $11$ or $00$. These rules have a simple interpretation~\cite{PRX17}: if the gate is applied across a bipartition boundary one has $I_k(t+1)=a I_{k-1}(t)+a I_{k+1}(t)$, whereas purities do not change, $I_k(t+1)=I_k(t)$, if the gate acts within one subsystem. Using this set of rules we can work backward from the last gate $M_{n-1,n}$ of the S configuration to the first $M_{1,2}$, and get all bipartitions that contribute to $I_k(t)$. For instance, taking $n=4$ and looking at the bipartition $1100$, i.e., $I_{k=2}$, the relevant bipartitions on previous steps are sketched in Fig.~\ref{fig:basis_example}. 
Reading the figure from top to bottom, we first look for purities (bipartitions) that contribute to $I_2(t)$ before the gate $M_{3,4}$ is applied. Looking at the rules from Eq.~\ref{eq:maps} we see that only $I_2$ can be mapped to $I_2(t)$ (we can reach $00$ on bits 3 and 4 only from $00$). Continuing by $M_{2,3}$, we now have two possibilities: we can obtain $I_2$ from either $I_1$ or $I_3$, in both cases with a prefactor $a$ (see Eq.~\ref{eq:Mij}). Lastly, see that $I_1$ is obtained after $M_{1,2}$ from $I_0$ and $I_2$, while $I_3$ can be obtained only from $I_3$ after application of $M_{1,2}$. Summarizing, one S iteration on $n=4$ qubits results in $I_2(t) = a^2 I_0(t-1) + a^2 I_2(t-1) + a I_3(t-1)$. Crucial is that at all steps only bipartitions with consecutive qubits in $A$ appear, i.e., only single-cut bipartitions $I_k$. With the above example we obtain one row of our reduced transfer matrix. To calculate the whole matrix, we have to repeat the same procedure for every bipartition that appears in the iteration (for the example with $n=4$ this means $0000$ and $1110$), until we end up with a closed subset of bipartitions ($0000$, $1100$, $1110$ and $1111$ for the example above). As a side remark, the closed subset that we obtain is the smallest subset ${\cal S}$ containing $1100$, which is invariant under the multiplication from the right with $M$; ${\cal S} M \subseteq {\cal S}$. Generalizing this procedure for arbitrary system size $n$ we learn that the subset is composed of all single-cut bipartitions, except for the $k=1$. With this we obtained an $n$ dimensional reduced transfer matrix, which can be used to propagate purities with the S OBC protocol for single-cut bipartitions. 

\begin{figure}[h]
    \begin{center}
        \includegraphics[width=70mm]{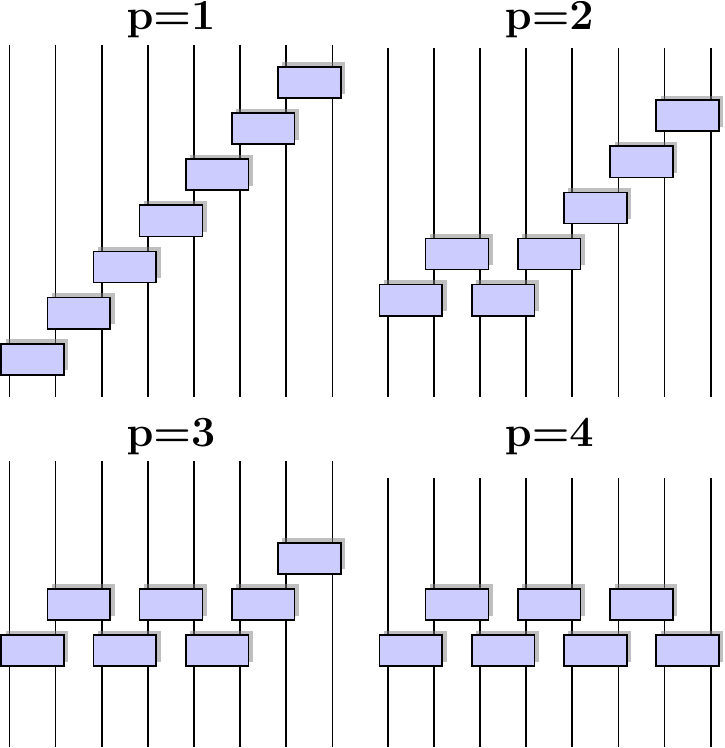}
        \caption{Canonical configurations for $n=8$. Different configurations are labeled with an integer $p=1,\ldots,n/2$, so that the BW section has $2p-1$ gates. For OBC (shown) one has in total $n-1$ gates, for PBC (not shown) an additional gate between qubits $n$ and $1$ is added at the end.}
        \label{fig:canonical}
    \end{center}
\end{figure}

In the following section we generalize this simplification to more complicated configurations. Namely, our analysis covers the dynamics of purity of an arbitrary bipartitions (not just a single-cut) and for a selection of $n/2$ OBC and $n/2$ PBC canonical protocols. By canonical protocols we mean random quantum circuits with gates ordered in a BW manner on the first $2p$ qubits and as an $S$ configuration for the rest \cite{foot1}, see Fig.~\ref{fig:canonical} for a depiction of all canonical protocols for $n=8$. The extreme cases are the S protocol for $p=1$, and the BW protocol for $p=n/2$.

\section{Single-cut bipartitions}
\label{sec:III}
\subsection{Transfer matrix}
\label{sec:IIIA}

Let us first consider canonical protocols and the simplest bipartition, namely one with a single-cut (automatically implying also OBC). For the S protocol we have seen that iteration by $M$ mixes only $I_k$, $k=0,2,\ldots,n-1,n$. Writing these single-cut bipartition purities into an $n$-dimensional vector $\mathbf{I}(t)=(1,I_2(t),I_3(t),\dots,I_{n-1}(t),1)$~\cite{foot1b}, where one can think of the first and last components representing trivial $I_0=I_n=1$, we can write the time evolution of average purities with a reduced transfer matrix as $\mathbf{I}(t+1)=A_{\mathrm{S}}^{(n)} \mathbf{I}(t)$ \cite{marko_22}, where
\begin{equation}
    A_{\mathrm{S}}^{(n)} = \begin{pmatrix}
        1 & 0 & 0 \\
        \textbf{a}_{\mathrm{S}} & S^{(n)}  & \textbf{b}_{\mathrm{S}} \\
        0 & 0 & 1
        \end{pmatrix},
\label{eq:S_red}
\end{equation}
and $\textbf{a}_{\mathrm{S}} = (a^2,a^3,\dots,a^{n-1})^T$, and $\textbf{b}_{\mathrm{S}} = (0,0,\dots,a)^T$ are $n-2$ dimensional column vectors, whereas $S^{(n)}$ is a $(n-2)\times(n-2)$ Toeplitz matrix,
\begin{align}
    S^{(n)}  &= \begin{pmatrix}
        a^2 & a & 0 & \dots & 0 \\
        a^3 & a^2 & a & \dots & 0 \\
        \vdots & \vdots & \ddots & \ddots & \vdots \\
        a^{n-2} & a^{n-3} & \ddots &\ddots & a \\
        a^{n-1} & a^{n-2} & \dots & a^3 & a^2
        \end{pmatrix}.
\label{eq:SR}
\end{align}
The superscripts $(n)$ of $A_{\mathrm{S}}^{(n)}$ and $S^{(n)}$ refer to the number of qubits (qudits).

The BW OBC protocol has a reduced form, too~\cite{marko_22}. The derivation of the reduced transfer matrix follows the same logic as for the S OBC case. If one wishes to study a single-cut bipartition with even $k$ in a system with even number of qubits, then one notices that the closed subspace consists of all single-cut bipartition purities with even qubits in $A$, i.e., $I_k$ with even $k$. The size of the subspace is $n/2+1$ and the vector containing purities that are propagated is $\mathbf{I}(t)=(1,I_2(t),I_4(t),\ldots,I_{n-2}(t),1)$. The reduced transfer matrix that evolves $\mathbf{I}(t)$ is
\begin{equation}
    A_{\mathrm{BW}}^{(n)}  = \begin{pmatrix}
        1 & 0 & 0 \\
        \textbf{a}_{\mathrm{BW}} & B^{(n)}  & \textbf{b}_{\mathrm{BW}} \\
        0 & 0 & 1
        \end{pmatrix},
\label{eq:BW_red}
\end{equation}
where $\textbf{a}_{\mathrm{BW}} = (a^2,0,\dots,0)^T$, $\textbf{b}_{\mathrm{BW}} = (0,0,\dots,a^2)^T$ and the tridiagonal matrix $B^{(n)}$ of size $(n/2-1)\times(n/2-1)$ is
\begin{align}
    B^{(n)}  &= a^2 \begin{pmatrix}
        2 & 1 &   \\
        1 & 2 & 1 &  \\
          & \ddots & \ddots & \ddots &  \\
          & & 1 & 2 & 1 \\
          & &  & 1 & 2 
        \end{pmatrix}.
\label{eq:BWR}
\end{align}
A crucial difference between the BW and S protocols is that $B^{(n)}$ is symmetric, causing purity decay to be given by $\lambda_2=4a^2$~\cite{adam18,Frank18,markov,AdamPRB19,u4_conjecture}, i.e., no surprises, while on the other hand for the S protocol $S^{(n)}$ is nonsymmetric, resulting in the decay not being given by $\lambda_2=4a^2$ (surprising phantom decay)~\cite{PRX,marko_22}.

We now generalize the reduced transfer matrix propagation to arbitrary OBC canonical protocols labeled by $p$. The procedure to obtain the reduced dynamics follows the same rules that were described for S OBC, so here we state only the result. Let us focus on the case where the cut is in the $S$ section, i.e., $k>2p$, or the cut is in the BW part and one has an even $k$ (for a qualitatively similar case of a cut in the BW section and odd $k$ see Appendix~\ref{app:oddk}). Purities for the relevant subset of bipartitions can be put into the vector $\mathbf{I}(t) = (1,I_2(t),I_4(t),\dots,I_{2p}(t),I_{2p+1}(t),I_{2p+2},\dots,1)$, which contains single-cut bipartitions with even $k$ for all $k \leq 2p$ (BW part of the canonical protocol) and all $I_k$ for $k\geq 2p$ (S part). The reduced transfer matrix that propagates $\mathbf{I}(t)$ turns out to be
\begin{equation}
    A_{p}^{(n)}  = 
    \begin{pmatrix}
        1 & 0 & 0 \\
        \textbf{a}_{\mathrm{BW}} & R_{p}^{(n)}  & \textbf{b}_{\mathrm{S}} \\
        0 & 0 & 1
    \end{pmatrix},\quad \mathbf{I}(t+1)=A_p^{(n)}\mathbf{I}(t),
\label{eq:p_red}
\end{equation}
with

\begin{equation}
    R_p^{(n)}  = 
    \left(\begin{array}{c|c}
        B^{(2p)} & \begin{array}{ccc}
                        0 & \dots & 0 \\
                        \vdots & \ddots & \vdots \\
                        a^2 & \dots & 0  
                  \end{array}   \\
                  \hline
        \begin{array}{cccc}
            0 & \dots & 0 &\mathbf{c}_p 
        \end{array}     & S^{(n-2p+2)}
    \end{array}\right),
\label{eq:pR}
\end{equation}
where $\mathbf{c}_p = (a^2, a^3, \dots, a^{n-2p+1})^T$, $S^{(n-2p+2)}$ is of dimension $(n-2p)\times(n-2p)$ and $B^{(2p)}$ of dimension $(p-1)\times(p-1)$. We see that the matrix $R_p^{(n)}$ contains an upper left block describing the BW part, and a lower right block for the S part. Both parts are ``coupled'' by the coefficients present in the vector $\mathbf{c}_p$ and a sole $a^2$ in the lower part of the upper right block. For example, for $n=14$ and $p=5$ we have $\mathbf{I}(t) = (1,I_2(t),I_4(t),I_6(t),I_8(t),I_{10}(t),I_{11}(t),I_{12}(t),I_{13}(t),1)$ and

\begin{equation}
    R_5^{(14)}  = 
    \begin{pmatrix}
         2 a^2 & a^2 & 0 & 0 & 0 & 0 & 0 & 0 \\
         a^2 & 2 a^2 & a^2 & 0 & 0 & 0 & 0 & 0 \\
         0 & a^2 & 2 a^2 & a^2 & 0 & 0 & 0 & 0 \\
         0 & 0 & a^2 & 2 a^2 & a^2 & 0 & 0 & 0 \\
         0 & 0 & 0 & a^2 & a^2 & a & 0 & 0 \\
         0 & 0 & 0 & a^3 & a^3 & a^2 & a & 0\\
         0 & 0 & 0 & a^4 & a^4 & a^3 & a^2 & a \\
         0 & 0 & 0 & a^5 & a^5 & a^4 & a^3 & a^2\\
   \end{pmatrix}.
\label{eq:pR_n16_p5}
\end{equation}
Using the propagation $\mathbf{I}(t+1)=A_p^{(n)} \mathbf{I}(t)$, the upper left $4 \times 4$ block of $R_5^{(14)}$ ``propagates'' $(I_2,I_4,I_6,I_8)$, while the lower right $4 \times 4$ is responsible for $(I_{10},I_{11},I_{12},I_{13})$.

\subsection{Spectral properties}
\label{sec:IIIB}

In \cite{PRX} it was proven that full transfer matrices $M$ for all possible permutations of $n-1$ gates under OBC (e.g., all canonical protocols $p$) share the same spectrum. Here we ask ourselves whether all \textit{reduced} transfer matrices $A_p^{(n)}$ also share the same spectrum. Namely, the spectrum of $A_p^{(n)}$ is only a subset of the spectrum of the nonreduced matrix $M$ and it could happen that the second largest eigenvalue of $A_p^{(n)}$ would be different (smaller) than the second largest eigenvalue of $M$. In Appendix \ref{app:spectrum} we prove that this is not the case. The nonzero part of the spectrum of $A_p^{(n)}$ is the same for all canonical protocols, specifically also $\lambda_2$, which is also equal to the second largest eigenvalue of the full $M$. Based solely on the spectrum the dynamics of purity would then be expected to decay as $|\lambda_2|^t$. We shall show that this is not the case and that purity asymptotically decays at a different rate.

More in detail, from the calculations found in Appendix~\ref{app:spectrum} we find that the nonzero eigenvalues $\lambda_j$ of $R_p^{(n)}$ are given by the zeros of the Chebyshev polynomial of the second kind
\begin{equation}
    U_{n/2-1}(1- \frac{\lambda_j}{2a^2}) = 0,
\label{eq:zeros} 
\end{equation}
that is 

\begin{equation}
    \lambda_j = 4a^2 \cos^2 \left( \frac{j\pi}{n} \right), \qquad j=1,\dots,n/2-1.
\label{eq:spectrum}
\end{equation}
The nonzero spectrum of $A^{(n)}_p$ is then obtained by adding to this a doubly degenerate eigenvalue 1. The kernel of $A_p^{(n)}$ is on the other hand a Jordan block of size $n/2-p$. The second largest eigenvalue of $A_p^{(n)}$ is the largest eigenvalue of $R_p^{(n)}$ and is therefore equal to $\lambda_2 = 4 a^2 \cos^2(\pi/n)$, is independent of $p$, and equal to the second largest eigenvalue of the nonreduced transfer matrix $M$.

\subsection{Effective purity decay $\leff$}
\label{sec:IIIC}

Here we shall compute the true asymptotic decay of purity in the thermodynamic limit for a single-cut bipartition, which is given by Eq.(\ref{eq:I_decay}) with 
\begin{equation}
  \leff=\frac{2}{3},\qquad k>2p,
  \label{eq:23}
\end{equation}
if the bipartition cut is in the S part of the configuration, while it is
\begin{equation}
  \leff=\lambda_2=\frac{16}{25},\qquad k<2p,
  \label{eq:1625}
\end{equation}
if the cut is in the BW part.

In Ref.~\cite{marko_22} three different methods were used to compute $\leff$ for S OBC circuits (i.e., $p=1$): (a) by finding the exact solution $I_k(t)$, (b) with the help of the spectrum of the operator in the TDL, and (c) by computing the pseudospectrum \cite{pseudo-spectrum} of the reduced transfer matrix. While those methods were feasible for the very simple structure of the S reduced matrix (\ref{eq:SR}), for our more complicated protocol there are difficulties: for (a) we are not able to get an exact solution, for method (b) it is not always clear how to define the operator in the TDL, e.g., in the case of transfer matrices should one take $p/n = \mathrm{const.}$ or $p = \mathrm{const.}$, etc., and, moreover, the computation of the operator spectrum could be challenging, (c) the pseudospectrum converges slowly with $n$. We therefore present a new method to compute $\leff$, which overcomes the above-mentioned difficulties.

The idea is to construct a new modified (perturbed) transfer matrix $\tilde{A}_p^{(n)}$ that obeys two conditions: (i) norms of left eigenvectors do not diverge in the TDL, and (ii) purity dynamics obtained by $A_p^{(n)}$ and the dynamics obtained by $\tilde{A}_p^{(n)}$ match up to times $t \sim n$. The first condition assures that the decay of purity is not slower than the decay given by the second largest eigenvalue $\tilde{\lambda}_2$ of $\tilde{A}_p^{(n)}$ after times $t_* \sim \ln n$ (see Appendix~\ref{app:proof_perturbation}). Thus it excludes a possible phantom decay with $\leff > \lambda_2$ at times $t \sim n$. In such a situation one would typically expect that the decay under $\tilde{A}_p^{(n)}$ will be correctly given by $\tilde{\lambda}_2^t$ already at nonextensive times larger than $t_1 \sim {\cal O}(1)$. In Appendix~\ref{app:lambda_eq_eff} we numerically check that this is indeed the case, except in cases where the initial vector is orthogonal to corresponding eigenvectors, a situation that happens for $k<2p$. In a nutshell, $\tilde{A}_p^{(n)}$ behaves essentially like a Hermitian matrix, even though it is not. Finally, the second condition (ii) then guarantees that the second largest eigenvalue of the perturbed matrix $\tilde{\lambda}_2$, which we calculate, will be equal to $\leff$ of the original unperturbed $A_p^{(n)}$ (for $k>2p$).

Let us construct such a matrix. We note that purity $I_{n-1}$ affects purity $I_k$ only after times $t \approx n-k$, so any change in the last row of $R_p^{(n)}$ will influence the dynamics of $I_k(t)$ only after $\sim n$ iterations, provided $k \neq n - \mathrm{const.}$. We define $\tilde{R}_p^{(n)}$ by changing the bottom leftmost value of the $S$-block in Eq.~\ref{eq:pR} from $a^{n-2p+1}$ to $a$, as can be seen below for $R_{5}^{(14)}$ 

\begin{equation}
    \tilde{R}_5^{(14)}  = 
    \begin{pmatrix}
         2 a^2 & a^2 & 0 & 0 & 0 & 0 & 0 & 0 \\
         a^2 & 2 a^2 & a^2 & 0 & 0 & 0 & 0 & 0 \\
         0 & a^2 & 2 a^2 & a^2 & 0 & 0 & 0 & 0 \\
         0 & 0 & a^2 & 2 a^2 & a^2 & 0 & 0 & 0 \\
         0 & 0 & 0 & a^2 & a^2 & a & 0 & 0 \\
         0 & 0 & 0 & a^3 & a^3 & a^2 & a & 0\\
         0 & 0 & 0 & a^4 & a^4 & a^3 & a^2 & a \\
         0 & 0 & 0 & a^5 & \textcolor{red}{a} & a^4 & a^3 & a^2\\
   \end{pmatrix}.
\label{eq:change_n16_p5}
\end{equation}
The change is marked in red. \new{The choice of such perturbation was made after trying different values at the bottom leftmost value of the $S$-block and taking the choice that makes the norm of left eigenvectors of the new transfer matrix converge in the TDL.} The new transfer matrix $\tilde{A}_p^{(n)}$ is defined as $A_p^{(n)}$ in Eq.~\ref{eq:p_red} by substituting $R_p^{(n)}$ with $\tilde{R}_p^{(n)}$. Remember that the largest eigenvalue of $\tilde{R}_p^{(n)}$ corresponds to the second largest eigenvalue of $\tilde{A}_p^{(n)}$.

Let us numerically calculate norms of left eigenvectors of $\tilde{R}_p^{(n)}$ in order to check whether they diverge. We normalize all right eigenvectors to $1$. Computing the largest norm of left eigenvectors, $\max_k || l_k || = \max_k \sqrt{ \langle l_k|l_k\rangle}$, for various system sizes $n$ and $p=n/5$ we observe that $\lim_{n \rightarrow \infty} \max_k || l_k || \approx 1$ \cite{foot2}, specifically it does not diverge with system size as is the case for $R_p^{(n)}$. Therefore both conditions (i) and (ii) are fulfilled for the new transfer matrix.

\begin{figure}[h]
    \begin{center}
        \includegraphics[width=85mm]{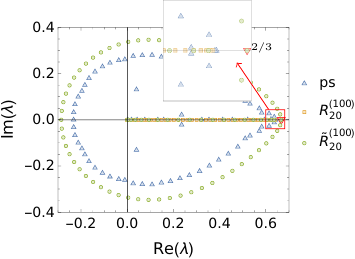}
        \caption{Calculating $\leff$ via a perturbed transfer matrix. Comparison between the spectrum of $R_{20}^{(100)}$, see Eq.~\ref{eq:pR}, (orange squares), $\tilde{R}_{20}^{(100)}$, obtained from Eq.~\ref{eq:tildeR_eigvals}, (green circles), and the pseudospectrum of $R_{20}^{(100)}$ (``ps", blue triangles). The largest pseudoeigenvalue is, despite large $\varepsilon=10^{-5}$, at $0.642$ still closer to the unperturbed eigenvalue $\lambda_2 \approx 0.639$, than to the correct $\leff=2/3$ (red triangle).}
        \label{fig:tildeR_eigvals}
    \end{center}
\end{figure}

We now compute the largest eigenvalue $\tilde{\lambda}_2$ of $\tilde{R}_p^{(n)}$. The derivation of the spectrum of $\tilde{R}_p^{(n)}$ follows almost the same steps as the procedure to calculate the spectrum of $R_p^{(n)}$ (see Appendix~\ref{app:spectrum}), so we omit the details and state only the result. The eigenvalues $\tilde{\lambda}$ of the new transfer matrix $\tilde{R}_p^{(n)}$ are solutions to the equation
\begin{align}
    \left( a^{n-2p} - 1 \right) & U_{2p-1}(\sqrt{\tilde{\lambda}}/2a) + \nonumber \\
    &(-1)^n \tilde{\lambda}^{n/2-p} U_{n-1}(\sqrt{\tilde{\lambda}}/2a) = 0.
\label{eq:tildeR_eigvals}
\end{align}
\new{We are not able to solve the equation above analytically, so we used standard numerical methods to find the roots of Eq.~\ref{eq:tildeR_eigvals}, i.e., the spectrum of $\tilde{R}_p^{(n)}$, which are plotted} in Fig.~\ref{fig:tildeR_eigvals} for $n=100$ and $p=20$. The largest eigenvalue of the modified matrix is $\tilde{\lambda}_2 \approx 2/3 - 3~\cdot~10^{-10}$, which is almost equal to $\leff = 2/3$, see red triangle in Fig.~\ref{fig:tildeR_eigvals}. The unperturbed matrix has on the other hand $\lambda_2 \approx 0.639$ (i.e., a finite-size value of $16/25=0.64$). As mentioned, another possibility to compute $\leff$ would be with the pseudospectrum of $R_p^{(n)}$. The pseudospectrum for matrices $A$ of finite size $n$ is approximated with the $\varepsilon$ pseudospectrum, $\mathrm{sp}_{\varepsilon}(A)$. The $\varepsilon$ pseudospectrum is defined as the spectrum of $A+\varepsilon E$, where $E$ is a matrix of univariate random Gaussian numbers. The pseudospectrum of $A$ is obtained by limiting $\lim_{n \rightarrow \infty} \mathrm{sp}_{\varepsilon}(A)$ while holding $\varepsilon$ fixed. Applications of pseudospectrum can be found in the context of skin effect \cite{Sato20,Sato21,Yoshida21} and metastability \cite{viola21}. From Fig.~\ref{fig:tildeR_eigvals} we see that using the pseudospectrum of $R_p^{(n)}$ as a proxy for $\leff$, like in Ref.~\cite{marko_22}, does not work well. Convergence of the $\varepsilon$ pseudospectrum with $n$ is slow and at $n=100$ the norm of the $\varepsilon$ pseudospectrum is still closer to $\lambda_2$ than to $\leff$. For example, to obtain $\leff$ within the absolute error $\pm 10^{-5}$ with the pseudospectrum one must take $n=20000$, meanwhile the spectrum of $\tilde{R}_p^{(n)}$ yields the same precision already with $n\approx 40$.

\begin{figure}[h]
    \begin{center}
        \includegraphics[width=70mm]{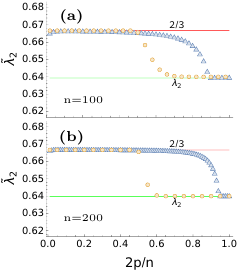}
        \caption{The largest eigenvalue of $\tilde{R}_p^{(n)}$ (blue triangles), compared with the actual purity decay $\leff$ computed numerically (orange circles), Eq.(\ref{eq:p_red}), all shown for $n=100$, $200$ and all possible $p$. The red and green line \new{(upper and lower horizontal line, respectively)} are set at $2/3$ and $\lambda_2 = 16/25$, which seems are the only two values that $\leff$ and $\tilde{\lambda}_2$ take in the TDL.}
        \label{fig:phantom_p}
    \end{center}
\end{figure}

We now compare the conjectured $\leff=\tilde{\lambda}_2=2/3$ with the actual purity decay of $I_{k=n/2}$ for arbitrary canonical protocol $p$. In Fig.~\ref{fig:phantom_p} we plot the exact effective rate $\leff$ obtained from full numerically calculated $I(t)$ (\ref{eq:p_red}) at large times, and $\tilde{\lambda}_2$, for all possible values of $p$. The largest eigenvalue of $\tilde{R}_p^{(n)}$ correctly predicts purity dynamics up to parameters $p \approx n/4$. For $p > n/4$, the actual decay is given by an eigenvalue of $\tilde{R}_p^{(n)}$ which is not the largest one. Comparing data for $n=100$ and $n=200$ suggests that in the TDL $\leff$ equals to $2/3$ when $p<n/4$ and to $16/25$ for $p>n/4$. Remembering that we are showing purity decay data for $k=n/2$ this corresponds exactly to a cut being either in the S, or the BW part of the circuit, respectively. Mathematically, the reason why $\tilde{\lambda}_2$ fails to predict the actual purity decay for $2p>k$ can be found by looking at the right eigenvectors of $\tilde{A}_p^{(n)}$. Purity can be expressed using the spectral decomposition of $\tilde{A}_p^{(n)}$ as 
\begin{equation}
    I_{k=n/2}(t) = \sum_j \tilde{\lambda}_j^t \langle e_{n/2}| r_j\rangle \langle l_j| (1,1,\dots,1) \rangle,
\label{eq:expansion}
\end{equation}
where $ \langle l_j|$ and $ |r_j\rangle$ are the left and right eigenvectors of $\tilde{A}_p^{(n)}$ and the vector $ \langle e_{n/2}|$ has a sole nonzero component at the position where $I_{n/2}(t)$ is found in $\mathbf{I}(t)$ (see the text above Eq.~\ref{eq:p_red}). Note that, for $p>n/4$, this position is $n/4$, otherwise it is greater than $n/4$ and maximally equal to $n/2$ for $p=1$. For every eigenvalue $\tilde{\lambda}_j$ of $\tilde{R}_p^{(n)}$ \new{of modulus} larger than $\approx 0.64$ all components $i<p$ of the corresponding right eigenvector are almost zero, which results in extremely small expansion coefficients $\langle e_{n/2}|r_j\rangle$ in Eq.~\ref{eq:expansion}. The largest contribution for $p>n/4$ thus comes from the eigenvalue $\tilde{\lambda}_j \approx 16/25$. Note that the discrepancy between $\tilde{\lambda}_2$ and $\leff$ in this case is not surprising, because it is due to the peculiarity of the initial condition and eigenvectors and could happen also in Hermitian systems. 

We have also looked at other single-cut bipartitions, i.e., $k \neq n/2$. The form of right eigenvectors of $\tilde{R}_p^{(n)}$ suggests that the actual purity decay in the TDL depends on $k$ and $p$: if $k > 2p$ then $\leff = 2/3$, otherwise $\leff = \lambda_2 = 16/25$. The condition $k>2p$ coincides with the fact that the boundary between the subsystem $A$ and $B$ lies in the S region of the canonical protocol; on the contrary, when $k<2p$ the boundary lies in the BW region. This gives the announced asymptotic purity decay with $\leff$ in Eqs.~(\ref{eq:23}) and (\ref{eq:1625}).

Equipped with understanding of single-cut bipartitions we next generalize this conjecture to bipartitions with multiple cuts, which also allows us to treat systems with PBC where one has at least two cuts.

\section{Multi-cut bipartitions}
\label{sec:IV}

Reduced transfer matrices for bipartitions that are not single-cut are more complicated and are constructed numerically. The algorithm to compute reduced transfer matrices is the same as the one that we used for canonical OBC protocols and single-cut bipartitions: we begin with the bipartition that we want to observe and obtain all purities at previous time that contribute to the wanted quantity. We continue recursively with the newly obtained bipartitions until we obtain a closed set of bipartitions $\alpha$. We numerically found that transfer matrix sizes depend on the number of boundaries $b$ between the subsystems $A$ and $B$ and scale as $\sim n^b$.

\begin{figure*}[]
    \begin{center}
        \includegraphics[width=150mm]{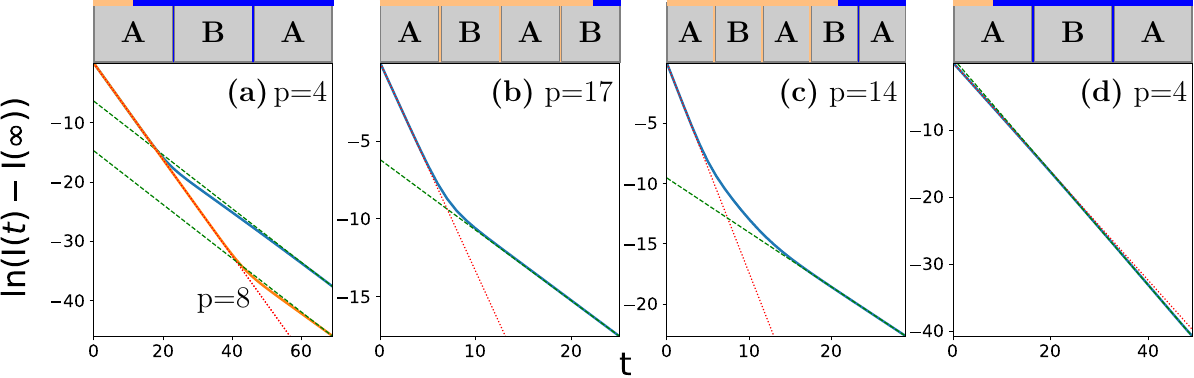}
        \caption{Purity evolution for different protocols $p$ and different bipartitions with multiple cuts. We use $n=40$ qubits bipartitioned into subsystems A and B as indicated above each figure, where also the thick \new{orange and blue (light and dark grey)} curves indicate the extend of the BW and S protocol sections, respectively. For instance, in panel (a) we have a canonical protocol with $p=4$ (blue curve), i.e., $8$ gates in BW are followed by $31$ gates in the S configuration, with the subsystem A containing first 14 and last 13 qubits. Panel (a) also shows data for $n=80$, $p=8$ (orange curve). Comparing data for $n=40$ and $n=80$ we find that the time until $\leff$ persist is $t \approx n/2$. Red dotted lines show the theoretical decay given by $\leff$, Eq.~\ref{eq:phantom_conjecture} ($\leff$ equal to $(2/3)^2, (16/25)^3, (16/25)^3(2/3)$ for panels (a),(b),(c), respectively), whereas green dashed lines are given by $\lambda_2$ of the reduced transfer matrix calculated by exact diagonalization. For panels (a)-(c) we use OBC so that $\lambda_2 \approx 0.64$, while for panel (d) we use PBC where $\lambda_2 \approx 0.430$ which is very close to $\leff=(2/3)^2$.}
        \label{fig:phantom_general}
    \end{center}
\end{figure*}

Figure~\ref{fig:phantom_general} shows purity dynamics obtained by numerically iterating the transfer matrix for various bipartitions and protocols. At each boundary between $A$ and $B$, we have two neighbouring qubits, one in $A$ and the other in $B$. On these qubits we act with either the S part or the BW part of the canonical protocol. We label the number of boundaries in the BW part with $\cbw$ and the number of boundaries under the S part with $\cs$. Based on numerical simulations (also other not shown) we conjecture that $\leff$ depends only on $\cbw$ and $\cs$. Namely,
\begin{equation}
    |I(t)-I(\infty)| \sim \leff^t, \qquad \leff = \left( \frac{16}{25} \right)^{\cbw} \left( \frac{2}{3} \right)^{\cs}.
\label{eq:phantom_conjecture}
\end{equation}
For cases where it is not clear whether the boundary lies in the BW or in the S part, see \cite{foot3}. The time until this conjecture holds, and when the decay $\lambda_2^t$ starts, is proportional to the length of the smallest contiguous region of qubits from one of the subsystems. Therefore, provided all contiguous parts composing A and B are extensive, the decay $\leff^t$ also holds up to an extensive time and therefore gives the relevant decay in the TDL. For example, see Fig.~\ref{fig:phantom_general}(a) for a comparison between two similar systems with $n=40$ and $n=80$. \new{Note that by changing the size of A or B parts in finite systems does not change the initial effective decay $\leff$ or the asymptotic decay $\lambda_2$. Changes in the sizes of A and B parts reflect themselves only in the time until $\leff$ persists.}

\begin{figure}[]
    \begin{center}
        \includegraphics[width=70mm]{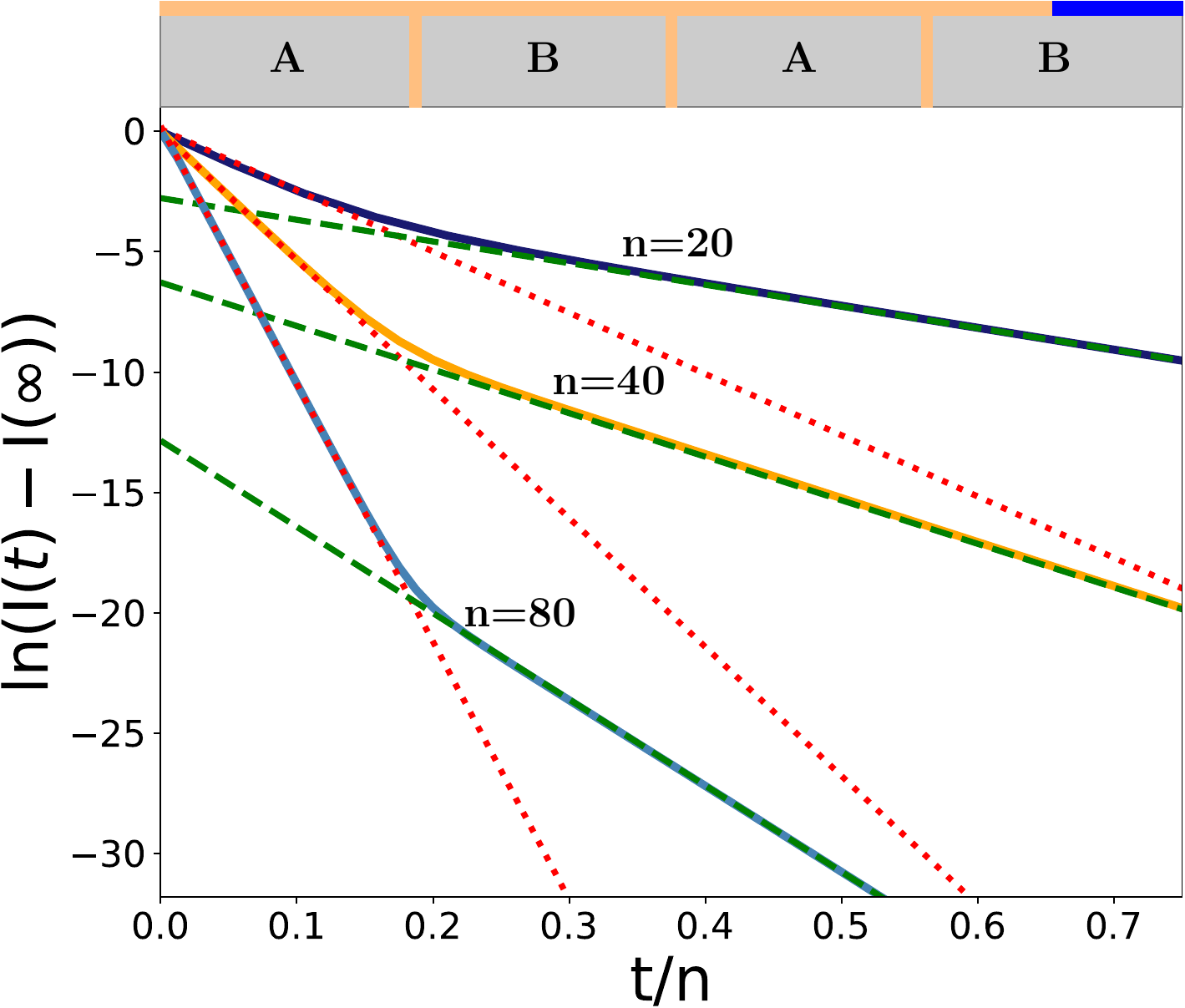}
        \caption{Purity decay for a bipartition composed of four regions (ABAB) of equal length for $n=20,40,80$, all for OBC, which demonstrates that, in the TDL, the transition from the rate given by $\leff$ to the rate given by $\lambda_2$ is discontinuous. The circuit geometries used are $p=9,17,34$ for $n=20,40,80$, respectively, so the bipartition always cuts the BW section three times and never cuts the S part (see the cartoon picture above the plot). All red and green dotted lines have slopes $\leff=(16/25)^3$ and $\lambda_2\approx 16/25$, respectively, even though the slopes seem different because of the scaled time $t/n$.}
        \label{fig:crossover}
    \end{center}
\end{figure}

\new{As can be seen from Fig.~\ref{fig:phantom_general}(c), there can be a substantial crossover time when the decay is neither $\leff^t$ nor $\lambda_2^t$. In Fig.~\ref{fig:crossover} we plot the time evolution of purity for circuits of same type for different system sizes to show that this crossover time vanishes in the TDL. In the limit of infinite systems we expect the decay rate of purity to have a discontinuous transition from $\leff$ to $\lambda_2$ at time scaling as $t\sim n$.}

We also numerically computed $\lambda_2$ of the reduced transfer matrices for the protocols and bipartitions studied as well as for noncanonical protocols with OBC and PBC. In all cases we found that the second largest eigenvalue always coincides with the second largest eigenvalue of the full nonreduced transfer matrix $M$. For OBC this $\lambda_2$ does not depend on the configuration and was conjectured\cite{u4_conjecture} to be $\lambda_2 = (4/5 \cos (\pi/n))^2$. In the TDL the case with $\cbw=1$ and $\cs=0$ is therefore the only protocol where $\leff = \lambda_2$. Moreover, $\cbw=0$ and $\cs=1$ is the only case in OBC where $\leff=2/3 > \lambda_2=16/25$, meaning that this is the only instance of a phantom eigenvalue where the decay is slower than any true eigenvalue of the transfer matrix. In PBC circuits $\lambda_2$ does depend on $p$~\cite{PRX} and the only known conjectures are for the BW protocol ($p=n/2$) where $\lambda_2 = (4/5 \cos (\pi/n))^4$, giving the TDL result $(16/25)^2$~\cite{foot4}, and for the S ($p=1$) where in the TDL $\lambda_2 = 4/9$. Different than for OBC, with PBC one has $\leff = \lambda_2$ for both S and BW protocols.

\section{Discussion}

In the paper we studied purity decay, i.e., bipartite entanglement growth, in random quantum circuits composed of two-site random Haar gates in a so-called canonical configuration that consists of an initial BW section acting on $2p$ qubits and a S part on the remaining $n-2p$ qubits. Average purities for different bipartitions can be propagated by a Markovian transfer matrix, which can be reduced down to a dimension that is only polynomial in the number of qubits. This allowed for some quasi exact results as well as for an efficient numerical study.

We observed that purity in finite systems decays in two stages: initially, up to times $t \sim n$, the exponential decay is $\leff^t$, afterwards it is given by the second largest eigenvalue of the transfer matrix $\lambda^t_2$. In the thermodynamic limit $n \to \infty$ the relevant decay is therefore given by $\leff$, and not $\lambda_2$, and furthermore, $\leff$ is in a generic situation different than $\lambda_2$. The decay is therefore not given by the Markovian gap, as one would naively expect. The reason lies in a nonsymmetric nature of the transfer matrix. \new{Based on results from \cite{PRX}, we expect other entanglement quantifiers, e.g., von Neumann entropy and Renyi entropy, to behave with a similar, two-step relaxation. However, a Markovian description is available only for purity.} Similar phenomena are expected also for other quantities like OTOC, as already observed in specific situations~\cite{PRR}, as well as in nonrandom systems.

In systems with open boundary conditions and a single boundary between subsystems the reduced transfer matrix is relatively simple. The value of $\leff$ depends on the location of the boundary: if it cuts across the staircase part of the protocol it is $\leff = 2/3$, while it is $\leff = \lambda_2 = 16/25$ if it cuts across the brick-wall part of the protocol. Therefore, if the cut is across the staircase part the decay of purity is slower than any true eigenvalue of the transfer matrix, i.e., it looks as if there would be a phantom ``eigenvalue'' $\leff=2/3$ in the spectrum. If a system has multiple boundaries between subsystems, as is, e.g., the case for periodic boundary conditions, the transfer matrix is more complicated and we studied it numerically. We conjecture that $\leff$ depends only on two numbers, the number of cuts or boundaries across the brick-wall and the number of cuts across the staircase sections of the protocol, provided all subparts are of extensive length. The usually discussed brick-wall circuit with a single-cut bipartition, where $\leff=\lambda_2$, is therefore special. \new{As a side remark, note that for OTOC the finite-size asymptotic decay is equal to purity's decay $\lambda_2$ \cite{PRR}; however, $\leff$ can be different.}

While mathematically the discrepancy between $\leff$ and $\lambda_2$ comes from the non-orthogonality between eigenvectors of the transfer matrix, the physical interpretation behind this phenomenon is still unclear. \new{One can argue that $\lambda_2$ cannot always be a physically relevant decay, because it can depend on boundary conditions, even though we expect purity decay to be independent of boundary conditions up to extensive times. A simple example can be found in BW circuits for OTOC, which exhibit a finite Lieb-Robinson velocity, which makes OTOC clearly insensitive to boundary conditions before times $t \sim n$ \cite{PRR}, $\lambda_2$ on the other hand is different for OBC and PBC.}

We would like to acknowledge support by Grants No.~J1-1698, J1-4385 and No.~P1-0402 from the Slovenian Research Agency, and MZ would like to thank also KITP and the National Science Foundation Grant No. NSF PHY-1748958. 

%\newpage

\onecolumngrid
\appendix
	
\section{Reduced transfer matrix for single-cut bipartitions with odd $k$ in the BW section}
\label{app:oddk}

In the main text, when constructing reduced transfer matrices for single-cut bipartitions and arbitrary protocols, we always constructed the subset of purities by beginning with a bipartition with an even number $k$ of qubits in the subsystem $A$. When $p\neq 1$, the resulting subset $S$ of purities, i.e., the basis in which $A_p^{(n)}$ is written, does not contain all single-cut bipartition purities. Namely, all purities $I_k$, $k$ odd and $k<2p$, are left out. In this appendix we explain how to change $A_p^{(n)}$ in order to propagate also these odd $k$ purities. For sake of clarity, we shall label purity $I_k$, $k$ odd and $k<2p$ with $I_{k_*}$.

First, we calculate all purities at time $t-1$ that contribute to $I_{k_*}$. Following the procedure explained in the main text we see that $I_{k_*} = a I_{k_*+1} + a I_{k_*-1}$. Note that both $I_{k_*+1}$ and $I_{k_*-1}$ are elements of $S$, so the new closed subset of purities will be $S' = S \cup \{ I_{k_*+1}\}$. We order $S'$ by increasing $k$ and denote the position of $I_{k_*}$ in this ordered set by $j$. Because $S'$ is obtained from $S$ just by adding the elements $I_{k_*}$, the new transfer matrix can be derived by adding a new row and a new column to $A_p^{(n)}$ at position $j$. The row that should be added to the transfer matrix is $(0,\dots,a,0,a,\dots,0)$, where the coefficients $a$ are on positions $j-1$ and $j$. How about the column? We already know that purities in the set $S$ do not depend on $I_{k_*}$, otherwise they would not form a closed set. This means that in order to accommodate $I_{k_{*}}$ we should add a column of zeros.

For instance, if we wish to propagate purity $I_5$ in a system with $n=14$ with the canonical protocol $p=5$, we propagate the vector $\mathbf{I} = (1,I_2,I_4,I_{5_*},I_6,I_8,I_{10},I_{11},I_{12},I_{13},1)$ with the transfer matrix $A_p^{(n)}$ from Eq.~\ref{eq:p_red}, where
\begin{equation}
    R_5^{(14)}  = 
    \begin{pmatrix}
         2 a^2 & a^2 & 0 & 0 & 0 & 0 & 0 & 0 & 0 \\
         a^2 & 2 a^2 & 0 & a^2 & 0 & 0 & 0 & 0 & 0 \\
         0 & a & 0 & a & 0 & 0 & 0 & 0 & 0 \\
         0 & a^2 & 0 & 2 a^2 & a^2 & 0 & 0 & 0 & 0 \\
         0 & 0 & 0 & a^2 & 2 a^2 & a^2 & 0 & 0 & 0 \\
         0 & 0 & 0 & 0 & a^2 & a^2 & a & 0 & 0 \\
         0 & 0 & 0 & 0 & a^3 & a^3 & a^2 & a & 0\\
         0 & 0 & 0 & 0 & a^4 & a^4 & a^3 & a^2 & a \\
         0 & 0 & 0 & 0 & a^5 & a^5 & a^4 & a^3 & a^2\\
   \end{pmatrix}.
\label{eq:pR_n16_p5}
\end{equation}

\section{Spectrum of the reduced transfer matrix for single-cut bipartitions and OBC}
\label{app:spectrum}

In this appendix we present a proof of the spectral equivalence of all reduced transfer matrices $A_p^{(n)}$. Let us denote with $\Sigma(C)$ the spectrum of a matrix $C$. First, we note that $\Sigma(A_p^{(n)}) = \Sigma(R_p^{(n)}) \cup \{1,1\}$, so that we shall focus on $\Sigma(R_p^{(n)})$ from now on. The largest eigenvalue in the spectrum of $R_p^{(n)}$ will be denoted as $\lambda_2$ and corresponds to the second largest eigenvalue of the reduced transfer matrix $A_p^{(n)}$. To compute the spectrum, we present a computation done for fixed $n$ and $p$, the generalization to general matrix size and canonical protocol will then be straightforward. That said, let us take $n=14$ and $p=5$. We wish to compute $\det(R_5^{(14)} - \lambda) = 0$, i.e.,

\begin{equation}
        \begin{vmatrix}
            2 a^2-\lambda  & a^2 & 0 & 0 & 0 & 0 & 0 & 0 \\
            a^2 & 2 a^2-\lambda  & a^2 & 0 & 0 & 0 & 0 & 0 \\
            0 & a^2 & 2 a^2-\lambda  & a^2 & 0 & 0 & 0 & 0 \\
            0 & 0 & a^2 & 2 a^2-\lambda  & a^2 & 0 & 0 & 0 \\
            0 & 0 & 0 & a^2 & a^2-\lambda  & a & 0 & 0 \\
            0 & 0 & 0 & a^3 & a^3 & a^2-\lambda  & a & 0 \\
            0 & 0 & 0 & a^4 & a^4 & a^3 & a^2-\lambda  & a \\
            0 & 0 & 0 & a^5 & a^5 & a^4 & a^3 & a^2-\lambda 
        \end{vmatrix} = 0.
\label{eq:det-R_p}
\end{equation}

We continue by using Laplace expansion around the fourth row (in general that would be the $(p-1)$st row). By using elementary properties of determinants (such as $\det(A \oplus B) = \det A \det B$) we get 

\begin{align}
    \det(R_5^{(14)} - \lambda) = 
    &-a^4     \begin{vmatrix} 
                2 a^2 - \lambda & a^2 \\
                a^2 & 2 a^2 - \lambda 
            \end{vmatrix}                   \begin{vmatrix}
                                                a^2-\lambda  & a & 0 & 0 \\
                                                a^3 & a^2-\lambda  & a & 0 \\
                                                a^4 & a^3 & a^2-\lambda  & a \\
                                                a^5 & a^4 & a^3 & a^2-\lambda  \\
                                            \end{vmatrix}                                          \nonumber \\
    &+(2 a^2 -\lambda)   \begin{vmatrix}
                            2 a^2-\lambda  & a^2 & 0 \\
                            a^2 & 2 a^2-\lambda  & a^2 \\
                            0 & a^2 & 2 a^2-\lambda  \\
                        \end{vmatrix}       \begin{vmatrix}
                                                a^2-\lambda  & a & 0 & 0 \\
                                                a^3 & a^2-\lambda  & a & 0 \\
                                                a^4 & a^3 & a^2-\lambda  & a \\
                                                a^5 & a^4 & a^3 & a^2-\lambda  \\
                                            \end{vmatrix}                                          \nonumber \\
    &-a^2               \begin{vmatrix}
                            2 a^2-\lambda  & a^2 & 0 \\
                            a^2 & 2 a^2-\lambda  & a^2 \\
                            0 & a^2 & 2 a^2-\lambda  \\
                        \end{vmatrix}       \begin{vmatrix}
                                                a^2 & a & 0 & 0 \\
                                                a^3 & a^2-\lambda  & a & 0 \\
                                                a^4 & a^3 & a^2-\lambda  & a \\
                                                a^5 & a^4 & a^3 & a^2-\lambda  \\
                                            \end{vmatrix}.
\label{eq:det-minors}
\end{align}
In the equation above we see a sum of products of two determinants. We recognize five out of these six determinants as determinant of either $\det(S^{(6)}-\lambda)$ (see Eq.~\ref{eq:SR}), $\det(B^{(6)}-\lambda)$ or $\det(B^{(8)}-\lambda)$ (Eq.~\ref{eq:BWR}). In the next step, we find a relation between the last determinant and $\det A_1^{(k)}$. Let us denote with $O^{(k)}$ determinants of the form

\begin{equation}
    \det O^{(k)} = \det (S^{(k)}-\lambda + \lambda E_{1,1}),
    \label{eq:det-defO}
    \end{equation}
where $E_{1,1}$ is a matrix having $1$ at position $(1,1)$ and $0$ elsewhere. We see that $\det O^{(6)}$ is present in the last term of Eq.~\ref{eq:det-minors}. Laplace expanding $\det O^{(k)}$ and $\det S^{(k)}$ around the first row we get two recursive relations

\begin{align}
    \det O^{(k)} &= a^2 \det S^{(k-1)} - a^2 \det O^{(k-1)} \\
    \det S^{(k)} &= (a^2 - \lambda) \det S^{(k-1)} - a^2 \det O^{(k-1)}.
\label{eq:det-recursion}
\end{align}

Combining the two recursions above we finally get $\det O^{(k)} = \det (S^{(k)} + \lambda S^{(k-1)})$. Plugging this in Eq.~\ref{eq:det-minors} we arrive at

\begin{align}
    \det(R_5^{(14)}-\lambda)    &= -a^4 \det(B^{(6)} - \lambda) \det(S^{(6)}-\lambda) \nonumber \\
                                & + (a^2-\lambda) \det(B^{(8)} - \lambda) \det(S^{(6)}-\lambda) \nonumber \\
                                &-a^2 \lambda \det(B^{(8)} - \lambda) \det(S^{(5)}-\lambda). 
\label{eq:det-SandBW}
\end{align}
The generalization from Eq.~\ref{eq:det-SandBW} to an arbitrary $n$ and $p$ is straightforward:

\begin{align}
    \det(R_p^{(n)}-\lambda) &= -a^4 \det(B^{(2p-2)} - \lambda) \det(S^{(n-2p+4)}-\lambda) \nonumber \\
                            & + (a^2-\lambda) \det(B^{(2p)} - \lambda) \det(S^{(n-2p+4)}-\lambda) \nonumber \\
                            &-a^2 \lambda \det(B^{(2p)} - \lambda) \det(S^{(n-2p+3)}-\lambda). 
\label{eq:gdet-SandBW}
\end{align}

In the next step, we link $\det(B^{(k)}-\lambda)$ to $\det(S^{(l)}-\lambda)$. An easy calculation shows that

\begin{equation}
    \det(B^{(k)} - \lambda) = a^{k-2} U_{k/2-1}(1-\lambda/(2a^2)),
\label{eq:gdet-BW}
\end{equation}
where $U_m(x)$ is the Chebyshev polynomial of the second kind, explicitly given by $U_m(\cos \phi) = \sin([(m+1)\phi]/\sin\phi)$. The characteristic polynomial $\det(S^{(k)}-\lambda)$ was found to be equal to $(-1)^k a^{k-1} \lambda^{(n-3)/2} U_{k-1}(\sqrt{\lambda} / 2a)$ \cite{eigs}. Using the fact that $U_{k/2-1}(1-x) = U_{k-1}(\sqrt{x/2}) (-1)^{k/2-1}/(2\sqrt{x/2})$ we arrive, after a rearrangement of the terms, at

\begin{align}
    \det(R_p^{(n)} -\lambda) = &(-1)^{n+3p} \lambda^{n/2-p-1} [ a^n U_{2p-5}(\sqrt{\lambda}/2a) U_{n-2p+1}(\sqrt{\lambda}/2a) + a^n U_{2p-3}(\sqrt{\lambda}/2a) U_{n-2p+1}(\sqrt{\lambda}/2a) - \nonumber \\
    &\lambda a^{n-2} U_{2p-3}(\sqrt{\lambda}/2a) U_{n-2p+1}(\sqrt{\lambda}/2a) + a^{n-1} \lambda^{1/2} U_{2p-3}(\sqrt{\lambda}/2a) U_{n-2p}(\sqrt{\lambda}/2a) ].
\label{eq:gdet-chebyshev}
\end{align}
The equation above can be simplified by considering the recursive relation for Chebyshev polynomials $U_{k+1}(x) = 2x U_{k}(x) - U_{k-1}(x)$

\begin{equation}
    \det(R_p^{(n)} -\lambda) = (-1)^{n+3p} \lambda^{n/2-p-1/2} a^{n-1} [ U_{2p-4}(\sqrt{\lambda}/2a) U_{n-2p+1}(\sqrt{\lambda}/2a) + U_{2p-3}(\sqrt{\lambda}/2a) U_{n-2p+2}(\sqrt{\lambda}/2a) ].
\label{eq:gdet-recursive}
\end{equation}
Finally, we exploit the relation $U_k(x) U_l(x) = \sum_{i=0}^l U_{k-l+2k}(x)$ (for $k\ge l$) to get

\begin{align}
    \det(R_p^{(n)} -\lambda) &= (-1)^{n+3p+1} \lambda^{n/2-p-1/2} a^{n-1} U_{n-1}(\sqrt{\lambda}/2a) \label{eq:gdet-final1}\\
                             &= (-1)^{n/2+3p} \lambda^{n/2-p} a^{n-2} U_{n/2-1}(1-\frac{\lambda}{2a^2}). \label{eq:gdet-final},
\end{align}
where we used the identity above Eq.~\ref{eq:gdet-chebyshev} to get from Eq.~\ref{eq:gdet-final1} to Eq.~\ref{eq:gdet-final}. The nonzero part of $\Sigma(R_p^{(n)})$ thus equals to zeros of the Chebyshev polynomial $U_{n/2-1}(1-\lambda/(2a^2))$. This polynomial is independent of $p$, so we conclude that reduced transfer matrices for all canonical protocols in OBC for single-cut bipartitions share the same nonzero part of the spectrum. The number of zero eigenvalues differs for different $p$, being the largest for $p=1$ and zero for $p=n/2$.

\section{The perturbed transfer matrix has no phantom eigenvalue}
\label{app:proof_perturbation}

In this appendix, we prove that a matrix $\tilde{A}_p^{(n)}$, of which left eigenvector norms converge in the TDL, has no phantom eigenvalue, i.e., the effective decay $\leff$ cannot be larger than the second largest eigenvalue $\tilde{\lambda}_2$ of $\tilde{A}_p^{(n)}$ after times $t>t_* \sim  \ln n$.

Suppose the largest eigenvalue of $\tilde{A}_p^{(n)}$ is equal to $\tilde{\lambda}_1 = 1$ and the second largest one is $\tilde{\lambda}_2 < 1$. We study the convergence of quantities $ \tilde{I}_j(t)= \langle e_j|(\tilde{A}_p^{(n)})^t| (1,1,\dots,1) \rangle $, $[e_j]_k = \delta_{j,k}$, to their asymptotic value $\tilde{I}_j(\infty)$. The quantity $\tilde{I}_j(t)$ was chosen to resemble purity from the main text (see Eq.~\ref{eq:expansion}).

The expression $\tilde{I}_j(t)$ can be rewritten in terms of the spectral decomposition of $\tilde{A}_p^{(n)} = \sum_k \tilde{\lambda}_k |r_k\rangle \langle l_k|$, where $\tilde{\lambda}_k$ are its eigenvalues and $ |r_k\rangle $ and $ \langle l_k|$ the corresponding right and left eigenvectors, respectively. We get

\begin{equation}
    \tilde{I}_j(t) = \sum_k \tilde{\lambda}_k^t \langle e_j|r_k\rangle \langle l_k|(1,1,\dots,1)\rangle. 
\label{eq:app_expansion}
\end{equation}
The asymptotic value $\tilde{I}_j(\infty)$ is equal to $ \langle e_j|r_1\rangle \langle l_1|(1,1,\dots,1)\rangle $. By using the triangle inequality and the fact that $|\tilde{\lambda}_2| \geq |\tilde{\lambda}_k|$, $k\neq 1$ we get an upper bound on the convergence rate

\begin{align}
    |\tilde{I}_j(t)-\tilde{I}_j(\infty)| &= |\sum_{k\neq1} \tilde{\lambda}_k^t \langle e_j|r_k\rangle \langle l_k|(1,1,\dots,1)\rangle| \nonumber \\
                       &\leq |\tilde{\lambda}_2|^t \sum_{k\neq1} |\langle e_j|r_k\rangle \langle l_k|(1,1,\dots,1)\rangle | = c_2 |\tilde{\lambda}_2|^t,
\label{eq:app_bound1}
\end{align}
where $c_2 = \sum_{k\neq1} |\langle e_j|r_k\rangle \langle l_k|(1,1,\dots,1)\rangle |$. In many cases studied in the main text, the decay rate towards the asymptotic value of purity differs from the one given by $\tilde{\lambda}_2$, instead one could have

\begin{equation}
    |\tilde{I}_j(t) - \tilde{I}_j(\infty)| \approx \ceff \leff^t,
\label{eq:app_leff}
\end{equation}
where $\leff > \tilde{\lambda}_2$. The effective decay $\leff$ cannot persist infinitely long, because at one point in time it will cross the upper bound from Eq.~\ref{eq:app_bound1} and we would get $\ceff \leff^{t_*} = c_2 |\tilde{\lambda}_2|^{t_*}$. If $t_*$ would scale as $\sim n$ or faster with $n$, then $\leff$ would be a phantom eigenvalue of $\tilde{A}_p^{(n)}$. We shall see that this is not possible.

Now, we present an argument for the claim $t_* \sim \ln n$. The time $t_*$ can be expressed as

\begin{equation}
    t_* = \frac{\ln(c_2/\ceff)}{\ln(\leff/|\tilde{\lambda}_2|)}.
\label{eq:app_tstar}
\end{equation}
The ratio $\leff/|\tilde{\lambda}_2|$ converges to a value $>1$ in the TDL, so it will not contribute to the dependence of $t_*$ on the system size. Next, we assume that $\ceff$ does not vanish in the TDL, so $\lim_{n \rightarrow \infty} \ceff = K > 0$. This assumption is motivated by the fact that we expect the effective decay $\leff$ to be present in the TDL. This argument shows that $\ln \ceff$ will not contribute to the $n$ dependence of $t_*$. The only contribution left is that of $\ln c_2$.

Next, we bound $\ln c_2$. If $\tilde{A}_p^{(n)}$ would be Hermitian, the left and right eigenvectors would be equal and could be normalized simultaneously; however, here we deal with a more general case, where $r_k \neq l_k$ and $ \langle l_k|r_j\rangle = \delta_{j,k}$. We can still normalize all right eigenvectors to $1$. Note that the left eigenvectors are not normalized and that their norms could be arbitrary large. Being right eigenvectors normalized, we get the bound $ |\langle e_j|r_k \rangle| \leq 1$, so 

\begin{equation}
    c_2 \leq \sum_{k \neq 1} | \langle e_j|r_k\rangle || \langle l_k|(1,1,\dots,1)\rangle| \leq \sum_{k\neq 1} |\langle l_k | (1,1,\dots,1) \rangle | \leq n \max_{k} \sum_j | [l_k]_j | \leq n^{3/2} \max_{k} ||l_k||,
\label{eq:app_bound2}
\end{equation}
where $||l_k|| = \sqrt{ \langle l_k|l_k\rangle}$ denotes the norm of the $k$-th left eigenvector. At the beginning of the appendix, we assumed that norms of left eigenvector of the matrix $\tilde{A}_p^{(n)}$ converge with $n$, so $\lim_{n \rightarrow \infty} \max_{k} ||l_k|| = C < \infty$. In turn, this means that $c_2 \sim n^{3/2}$ and

\begin{equation}
    t_* \sim \ln n.
\label{eq:app_final}
\end{equation}
Note that the obtained bound on $t_*$ will not be of much help if we iterate matrices of size exponential in $n$ as in \cite{PRX,PRR}. However, in this paper we deal with matrices of size polynomial in $n$.

\section{Emergence of the asymptotic decay with $\tilde{\lambda}_2$}
\label{app:lambda_eq_eff}

\begin{figure}[h]
    \begin{center}
    \includegraphics[width=150mm]{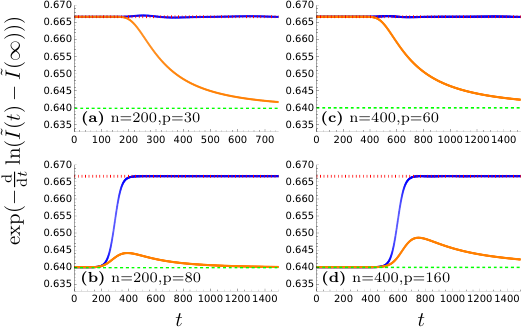}
        \caption{Determining $t_1$ when the decay of $\tilde{I}_{n/2}$ becomes $\tilde{\lambda}_2^t$. The instantaneous decay rate of $\tilde{I}_{n/2}$ \new{(blue/dark curves)} is compared with that of $I_{n/2}$ \new{(orange/light curves)}, seeing that the two agree up to times $\sim n$. Furthermore, in panels (a) and (c) where $2p<k$ this decay rate is immediately equal to $\tilde{\lambda}_2 = 2/3$ (red dotted line) and therefore $t_1 \approx 1$. On the other hand in panels (b) and (d) where $2p>k$, it is equal to $\lambda_2 = 16/25$ (green dashed line) up to times $\approx 3n/2$, so that $t_1$ in the TDL is formally infinite due to the initial state being orthogonal to respective eigenvectors.}%
        \label{fig:perturbed_iter}
        \end{center}
        \end{figure}

We know that for any finite size at a sufficiently large time $t_1$, $\tilde{I}_{n/2}(t)$ will eventually decay as $\sim \tilde{\lambda}_2^t$. In this appendix we are interested in the value of this $t_1$ when such asymptotic decay kicks in. Recall that $\tilde{I}_{n/2}(t)$ is defined as $\tilde{I}_{n/2}(t)= \langle e_{n/2}|(\tilde{A}_p^{(n)})^t| (1,1,\dots,1) \rangle$ (see Appendix~\ref{app:proof_perturbation}), where $\tilde{A}_p^{(n)}$ is the perturbed transfer matrix from the main text. Quantities $\tilde{I}(t)$, obtained numerically with iterations of $\tilde{A}_p^{(n)}$, are represented in Fig.~\ref{fig:perturbed_iter}. As expected, the dynamics of $\tilde{I}(t)$ and $I(t)$ match up to extensive times. Generalizing the results from Fig.~\ref{fig:perturbed_iter} (see also Fig.~\ref{fig:phantom_p}), when $p<n/4$ one has $\leff$ equals the second largest eigenvalue $\tilde{\lambda}_2$ immediately, meaning $t_1 = 1$. On the other hand, when $p>n/4$ $\leff = \tilde{\lambda}_2$ only after times $t_1 \approx 3n/2$, meaning that $t_1$ diverges with $n$. In this case computing the second largest eigenvalue of the perturbed transfer matrix does not help us to get the effective decay $\leff$ of purities $I_{n/2}$. In the main text we give a mathematical explanation for why in this case the finite-size asymptotic decay kicks in so late.

\begin{thebibliography}{99}
\bibitem{foot0} While in our cases there are no Jordan blocks at any finite $n$ (except for the kernel, whose effect though is not crucial), one can perhaps think of the phenomenon as arising due to an emerging effective Jordan block in the thermodynamic limit.
\bibitem{foot1} The motivation for focusing on canonical protocols comes from \cite{PRX}, where a spectral equivalence was proven. Namely, for PBC spectra of $M$ all factorially many different protocols can be grouped into just $n/2$ spectral equivalence classes, each having a representative canonical protocol. For OBC it is even simpler -- there is only one equivalence class. Therefore, for OBC all canonical protocols have exactly the same spectrum of $M$, including $\lambda_2$, while, as we will see, purity decay rates can be very different.
\bibitem{foot1b} Beware that $\mathbf{I}(t)$ now contains only purities for single-cut bipartitions $I_k(t)$, contrary to Section~\ref{sec:IIA}, where $\vec{I}(t)$ was used to label the vector containing purities for all $2^n$ bipartitions.
\bibitem{foot3} When the bipartition cuts exactly between the S and BW part of the canonical protocol it is not clear whether to increment $\cs$ or $\cbw$. In order to explore one such example we studied the PBC protocol and a bipartition composed of two sections $A$ and $B$. There are two boundaries between the subsystems of the bipartition: one in the bulk of the circuit and one between the first and the last qubit. The first cut contributes to either $\cs$ or $\cbw$, depending on the length of $A$. However, it is not clear how the second cut influences the purity decay rate. By looking at various system sizes ($n=40,60,100$) and bipartition lengths $|A|=n/3,2n/3$ we conclude that the second cut increases $\cbw$ by one, so it must be considered as a cut in the BW section of the protocol. Interestingly, when the bipartition cuts between the BW and the S part in the bulk of the circuits it is $\cs$ which must be incremented (based on numerical data for \{$n=100,200$, $|A|=n/2$, $p=n/4$\}, \{$n=90$, $|A|=n/3$, $p=n/6$\} and \{$n=90,180$, $|A|=2n/3$, $p=n/3$\}).
\bibitem{foot4} For BW with PBC and even $n$ one has in the TDL an exact expression $I(t)=(4/5)^{4t}$ if $n$ is an integer multiple of $4$, and $I(t)=(5/4)(4/5)^{4t}$ otherwise.
\end{thebibliography}
\end{document}